\documentclass[twocolumn]{webofc}
\usepackage[varg]{txfonts}   
\usepackage{graphicx}
\newcommand{\ba}{\begin{eqnarray}}
\newcommand{\ea}{\end{eqnarray}}
\newcommand{\bsub}{\begin{subequations}}
\newcommand{\esub}{\end{subequations}}

\def\ket#1{|#1\rangle}

\begin{document}
%
\title{Partial dynamical symmetry and the vibrational structure 
of Cd isotopes}
%
%

\author{\firstname{A.} \lastname{Leviatan}\inst{1}\fnsep
\thanks{\email{ami@phys.huji.ac.il}} \and
        \firstname{N.} \lastname{Gavrielov}\inst{1}\fnsep
\thanks{\email{noam.gavrielov@mail.huji.ac.il}} \and
        \firstname{J.E.} \lastname{Garc\'\i a-Ramos}\inst{2}\fnsep
\thanks{\email{enrique.ramos@dfaie.uhu.es}} \and
        \firstname{P.} \lastname{Van Isacker}\inst{3}\fnsep 
\thanks{\email{isacker@ganil.fr}}
}

\institute{Racah Institute of Physics, The Hebrew University, 
Jerusalem 91904, Israel
\and
Departamento de Ciencias Integradas, Universidad de Huelva, 21071 
Huelva, Spain
\and
Grand Acc\'el\'erateur National d'Ions Lourds, CEA/DRF-CNRS/IN2P3,
Bvd Henri Becquerel, B.P.~55027, F-14076 Caen, France
    }

\abstract{%
The recently reported deviations of selected non-yrast states in $^{110}$Cd 
from the expected spherical-vibrator behaviour, is addressed by means 
of an Hamiltonian with U(5) partial dynamical symmetry. The latter preserves 
the U(5) symmetry in a segment of the spectrum and breaks it in other states. 
The effect of intruder states is treated in the framework of the 
interacting boson model with configuration mixing.
}
\maketitle
%
The Cd isotopes 
have been traditionally considered to be a prime example of spherical 
vibrators. Recently, advanced experimental studies have reported 
significant deviations from this behaviour in 
selected two- and three-phonon states, along the 
Cd chain (A=108-126)~\cite{Garrett08,Garrett12,Batch14}. 
These  observations have 
led to claims for the ``breakdown of the vibrational motion" in these 
isotopes and the need for a paradigm shift~\cite{Garrett08,Garrett12}. 
In the present contribution, we examine an alternative explanation
for the structure of the Cd isotopes, in terms of U(5) partial dynamical
symmetry (PDS)~\cite{Leviatan11}.

A convenient starting point for describing spherical nuclei is the 
U(5) limit of the interacting boson model 
(IBM)~\cite{IBM}, corresponding to the chain of nested algebras,
\ba
{\rm U(6)\supset U(5)\supset SO(5)\supset SO(3)} ~. 
\label{U5-DS}
\ea
The basis states $\ket{[N],n_d,\tau,n_{\Delta},L}$ 
have quantum numbers which are the labels of irreducible 
representations of the algebras in the chain. 
Here $N$ is the total number of monopole ($s$) and quadrupole ($d$) 
bosons, $n_d$ and $\tau$ are the $d$-boson number and seniority, 
respectively, and $L$ is the angular momentum. 
The multiplicity label $n_{\Delta}$ counts the 
maximum number of $d$-boson triplets coupled to $L\!=\!0$. 
The dynamical symmetry (DS) Hamiltonian has the form
\ba
\hat{H}_{\rm DS} = 
t_1\,\hat{n}_d + t_2\,\hat{n}_{d}^2
+ t_3\,\hat{C}_{{\rm SO(5)}} + t_4\,\hat{C}_{{\rm SO(3)}} ~,
\label{H-DS}
\ea
where $\hat{C}_{\rm G}$ is the Casimir operator of G, and 
$\hat{n}_d\!=\!\sum_{m}d^{\dag}_md_m\!=\!\hat{C}_{{\rm U(5)}}$. 
$\hat{H}_{\rm DS}$ is completely 
solvable with eigenstates $\ket{[N],n_d,\tau,n_{\Delta},L}$ and 
eigen-energies
\ba
E_{\rm DS} =
t_1\, n_d + t_2\, n_{d}^2 
+ t_3\, \tau(\tau+3) + t_4\, L(L+1) ~.
\label{E-DS}
\ea
A typical U(5)-DS spectrum exhibits $n_d$-multiplets of 
a spherical vibrator, with a two-phonon ($n_d\!=\!2$) triplet of 
states ($L\!=\!4,2,0$) at an energy 
$E(n_d\!=\!2)\!\approx\! 2E(n_d\!=\!1)$ 
above the ground state ($n_d\!=\!L\!=\!0$),
and a three-phonon ($n_d\!=\!3$) quintuplet of states 
($L\!=\!6,4,3,0,2$) at $E(n_d\!=\!3)\approx 3E(n_d\!=\!1)$.
A quadrupole operator proportional to
\ba
\hat{Q} = d^{\dag}s + s^{\dag}\tilde{d} ~,
\label{Q}
\ea
enforces strong ($n_d+1\!\rightarrow\! n_d$) E2 transitions 
with particular ratios, {\it e.g.}, 
$\frac{B(E2;\,n_d =2, L=0,2,4\rightarrow n_d=1,L=2)}
{B(E2;\,n_d=1,L=2\rightarrow n_d=0,L=0)}
=2\frac{(N-1)}{N}$.
\begin{figure*}
\centering
\includegraphics[width=\linewidth,clip]{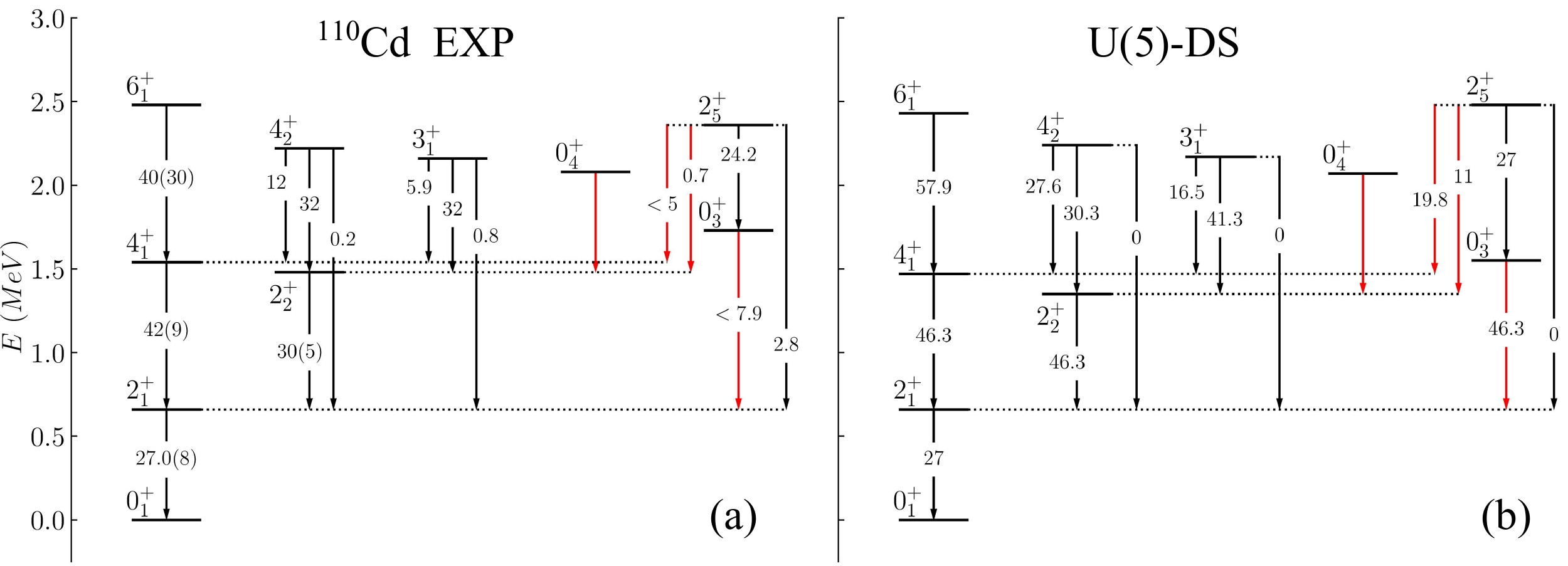}
\caption{
(a)~Experimental spectrum and representative 
E2 rates~\cite{Garrett12,NDS12} (in W.u.)  of normal levels 
in $^{110}$Cd. 
(b)~U(5)-DS spectrum obtained from $\hat{H}_{\rm DS}$~(\ref{H-DS}) with 
parameters 
$t_1\!=\!641.5,\, t_2\!=\!67.9,\, t_3\!=\!-26.1,\, 
t_4\!=\!8.7$ keV and $N\!=\!7$. 
The E2 operator is $e_{B}\,\hat{Q}$, Eq.~(\ref{Q}), with 
$e_{B}\!=\!1.96$ $(\rm W.u.)^{1/2}$. 
For additional experimental error bars on B(E2) values, not shown 
in Figs.~1(a) and 2~(a), see~\cite{Garrett12,NDS12}.}
\label{fig-1}
\end{figure*}
\begin{figure*}
\centering
\includegraphics[width=\linewidth,clip]{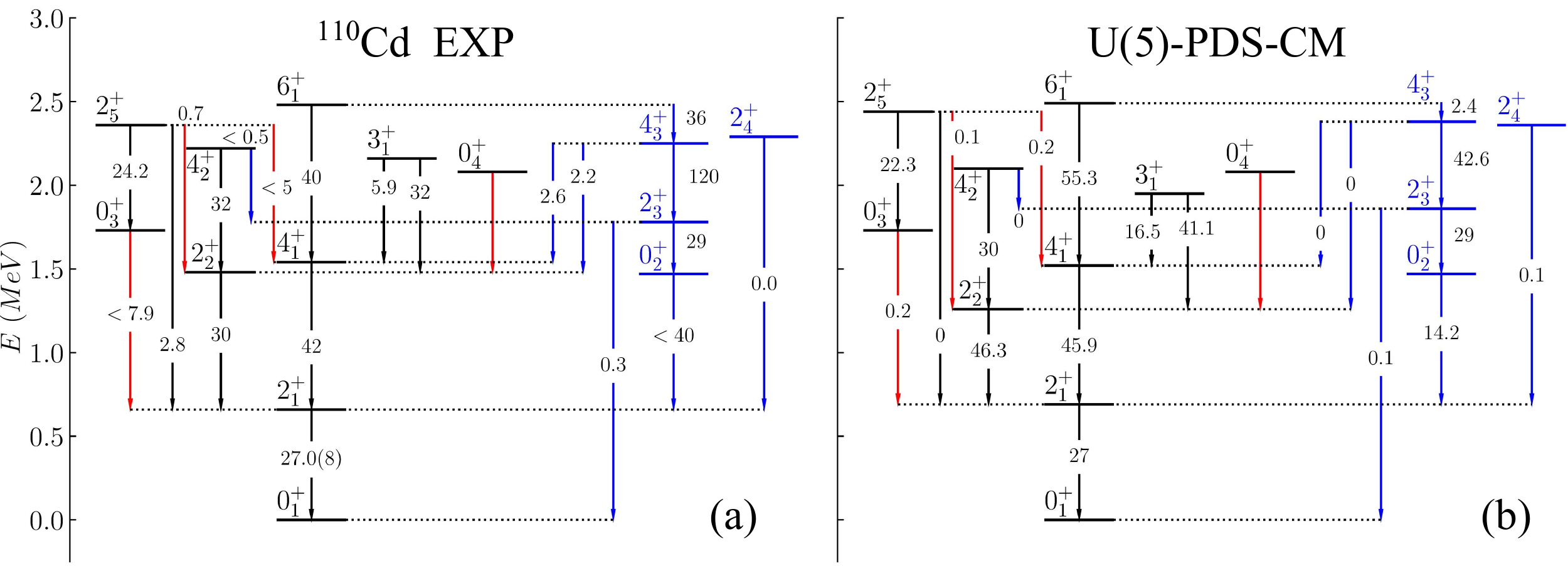}
\caption{
(a)~Experimental spectrum and representative 
E2 rates~\cite{Garrett12,NDS12} (in W.u.) of normal [as in Fig.~1(a)] 
and intruder levels ($0^{+}_2,\,2^{+}_3,\,4^{+}_3,\,2^{+}_4$) 
in $^{110}$Cd. (b)~U(5)-PDS-CM spectrum, 
obtained from $\hat{H}$~(\ref{Hfull}) with 
parameters 
$t_1\!=\!767.8,\, t_2\!=\!-t_3\!=\!73.6,\, 
t_4\!=\!18.5,\,r_0=2.1,\, e_0\!=\!-6.9,\,
\kappa\!=\!-72.7,\,\Delta/2=4989.5,\,\alpha\!=\!-42.8$ 
keV and $N\!=\!7$.
The E2 operator~(\ref{T-E2}) used has 
$e_{B}^{(N)}\!=\!1.96$, $e_{B}^{(N+2)}\!=\!1.19$ $(\rm W.u.)^{1/2}$.}
\label{fig-2}
\end{figure*}

The empirical spectrum of $^{110}$Cd consists of both 
normal levels [shown in Fig.~1(a)], 
and intruder levels [shown in Fig.~2(a)] based 
on 2p-4h proton excitations across the Z=50 closed shell. 
The experimental energies and E2 rates in Fig.~1(a), demonstrate 
that most normal states have good spherical vibrator properties, 
and conform well with the U(5)-DS calculation shown in Fig.~1(b). 
However, the measured 
rates for E2 decays from the non-yrast states, $0^{+}_3,\,(n_d=2)$ 
and $[0^{+}_4,\, 2^{+}_5\,(n_d=3)]$, 
reveal marked deviations from this behaviour. 
In particular, 
$B(E2;\,0^{+}_3\!\rightarrow\! 2^{+}_1) \!<\! 7.9$, 
$B(E2;\,2^{+}_5\!\rightarrow\! 4^{+}_1) \!<\! 5$,
$B(E2;\,2^{+}_5\!\rightarrow\! 2^{+}_2) \!<\! 0.7^{+0.5}_{-0.6}$ W.u.,  
are extremely small compared to the 
U(5)-DS values: 46.29, 11.02, 19.84 W.u., respectively. 
Absolute B(E2) values for transitions from the $0^{+}_4$ state 
are not known, but its branching ratio to $2^{+}_2$ is small.

Attempts to explain the above deviations in terms of strong 
mixing between the normal spherical [U(5)-like] states and 
intruder deformed [SO(6)-like] states have been shown to be 
unsatisfactory~\cite{Garrett08,Garrett12}. 
The reasons are two-fold. (i)~The strong mixing required for 
an adequate description of the two-phonon $0^{+}_3$ state, 
results in serious disagreements with the observed 
decay pattern of three-phonon yrast states. 
(ii)~The discrepancy in the decays of the non-yrast two- 
and three-phonon states 
persists also in the heavier $^{\rm A}$Cd isotopes (A=110-126), 
even though the energy of intruder states rises 
away from neutron mid-shell, and the mixing is reduced.
These observations have led to the conclusion that
the normal-intruder strong-mixing scenario needs to be rejected, 
and have raised serious questions on the validity of the 
multi-phonon interpretation~\cite{Garrett08,Garrett12}. 
In what follows, we consider a possible explanation 
for the ``Cd problem'', based on a partial dynamical symmetry (PDS).
The latter corresponds to a situation in which a given symmetry is obeyed by 
only a subset of states and is broken in other states. Such a notion 
has been previously employed in nuclear spectroscopy in conjunction with 
the SU(3)-DS~\cite{Leviatan96,LevSin99,levramisa13} and 
SO(6)-DS~\cite{Isacker99,LevIsa02,Ramos09} 
chains of the IBM. 
In the present contribution, we show the relevance of 
U(5)-PDS to the Cd problem.

The lowest normal levels comprise three classes of states,
\bsub
\ba
{\rm Class\,A:}&& n_d=\tau=\!0,1,2,3\quad (n_{\Delta} = 0)~,
\label{ClassA}\\
{\rm Class\,B:}&& n_d = \tau+2=2,3\,\quad (n_{\Delta} = 0)~,
\label{ClassB}\\
{\rm Class~C:}&& n_d = \tau = 3\,\;\;\quad\qquad (n_{\Delta} = 1)~. 
\label{ClassC}
\ea
\label{ClassABC}
\esub
In the U(5)-DS calculation of Fig~1(b), 
the ``problematic'' states 
$[0^{+}_3\,(n_d\!=\!2)$ and $2^{+}_5\,(n_d\!=\!3)]$  
belong to class~B, and $0^{+}_4\, (n_d\!=\!3)$ belongs to Class~C.
The remaining ``good'' spherical-vibrator states 
$[0^{+}_1\,(n_d\!=\!0);\,2^{+}_1\,(n_d\!=\!1);\,
4^{+}_1,2^{+}_2\,(n_d\!=\!2);\,6^{+}_1,4^{+}_2,3^{+}_1\,(n_d\!=\!3)]$ 
belong to Class~A. As mentioned, the spherical-vibrator interpretation 
is valid for most states in Fig.~1(a), but not all. 
We are thus confronted with a situation in which some states 
in the spectrum (assigned to Class~A) obey the predictions of U(5)-DS, 
while other states (assigned to Classes B and C) do not. 
These empirical findings signal the presence of a partial dynamical 
symmetry, U(5)-PDS.

The construction of Hamiltonians with U(5)-PDS follows the general 
algorithm~\cite{Ramos09,Alhassid92},
by identifying operators which annihilate particular sets of 
U(5) basis states. In the present case, we consider the following 
interaction,
\ba
\hat{V}_0 &=& r_0\,G^{\dag}_{0}G_{0}
+ e_{0}\,\left (G^{\dag}_0 K_0 + K^{\dag}_{0}G_0 \right ) ~,
\label{V0}
\ea
where $\textstyle{G^{\dag}_{0} \!=\! [(d^\dag d^\dag)^{(2)} d^\dag]^{(0)}}$, 
$K^{\dag}_{0} \!=\! s^{\dag}(d^{\dag} d^{\dag})^{(0)}$ 
and standard notation of angular momentum coupling is used. 
$\hat{V}_0$ of Eq.~(\ref{V0}) is in normal-ordered form and satisfies
\ba
\hat{V}_0\vert [N], n_d=\tau, \tau, n_{\Delta}=0, L \rangle = 0 ~,
\label{V0vanish}
\ea
with $L\!=\!\tau,\tau+1,\ldots,2\tau-2,2\tau$.
Eq.~(\ref{V0vanish}) follows from the fact
that the indicated states have $n_d\!=\!\tau$ and $n_{\Delta}=0$, hence 
do not contain a pair or a triplet of $d$-bosons 
coupled to $L=0$ and, as such, are annihilated by 
$K_0$~\cite{IBM} and $G_0$~\cite{Talmi03}.

The states of Eq.~(\ref{V0vanish}), which include those of Class~A,
form a subset of U(5) basis states, hence remain solvable eigenstates 
of the following Hamiltonian
\ba
\hat{H}_{\rm PDS} &=& \hat{H}_{\rm DS} + \hat{V}_0 ~,
\label{H-PDS}
\ea
with good U(5) symmetry and energies given in 
Eq.~(\ref{E-DS}) with $n_d=\tau$. 
It should be noted that while $\hat{H}_{\rm DS}$~(\ref{H-DS}) is diagonal 
in the U(5)-DS chain~(\ref{U5-DS}), 
the $r_0$-term ($e_0$-term) in $\hat{V}_0$ connects states with 
$\Delta n_d\!=\!0$ and $\Delta\tau=0,\pm 2,\pm 4,\pm 6$ 
$(\Delta n_d=\pm 1$ and $\Delta\tau=\pm 1,\pm 3)$.
Accordingly, the remaining eigenstates of $\hat{H}_{\rm PDS}$~(\ref{H-PDS}), 
in particular those of classes B and C, 
are mixed with respect to U(5) and SO(5). 
The U(5)-DS is therefore preserved in a subset of eigenstates 
but is broken in other states. 
By definition, $\hat{H}_{\rm PDS}$ exhibits U(5)-PDS. 
Cubic terms of the type present in $\hat{V}_0$, Eq.~(\ref{V0}),
are frequently encountered in PDS Hamiltonians, 
{\it e.g.}, in conjunction with 
signature splitting~\cite{levramisa13},
band structure~\cite{Isacker99,LevIsa02,Ramos09}, 
and shape-coexistence~\cite{LevDek16,LevGav17} in nuclei. 

The effect of intruder levels can be studied in the framework of the 
interacting boson model with configuration mixing (IBM-CM)~\cite{DuvBar82}.
The latter involves the space of normal states described by a 
system of $N$ bosons representing valence nucleon pairs,  
and the space of intruder states described by a system of $N\!+\!2$ bosons, 
accounting for particle-hole shell model excitations. 
This procedure has been used extensively in describing 
coexistence phenomena in 
nuclei~\cite{heyde95,Foisson03,ramos11,ramos14}. 
In the present study of $^{110}$Cd, 
the Hamiltonian in the normal sector is taken to be $\hat{H}_{\rm PDS}$ of 
Eq.~(\ref{H-PDS}), acting in a space of $N=7$ bosons. 
The Hamiltonian in the intruder sector is taken to be 
of SO(6)-type~\cite{heyde95},
\ba
\hat H_{\rm intrud} &=& \kappa \hat{Q}\cdot \hat{Q} + \Delta ~,
\label{H-intrud}
\ea
acting in a space of $N=9$ bosons, with $\hat{Q}$ given in Eq.~(\ref{Q}). 
A~mixing term between the $[N]$ and $[N\!+\!2]$ boson spaces is defined 
as~\cite{heyde95,Foisson03,ramos11,ramos14},
\ba
\hat V_{\rm mix} =  
\alpha \left [(s^{\dagger})^{2} + (d^{\dagger}d^{\dagger})^{(0)}\right ] 
+ {\rm H.c},
\label{Vmix}
\ea
where H.c. means Hermitian conjugate. The combined Hamiltonian for the 
two configurations has the form
\ba
\hat{H} = \hat{H}_{\rm PDS}^{(N)} + \hat{H}_{\rm intrud}^{(N+2)} 
+ \hat {V}_{\rm mix}^{(N,N+2)} ~.
\label{Hfull}
\ea
Here $\hat{\cal O}^{(N)} \!=\! \hat{P}^{\dag}_{N}\,\hat{\cal O}\,\hat{P}_{N}$ 
and $\hat{\cal O}^{(N,N')} \!=\! \hat{P}^{\dag}_{N}\,\hat{\cal O}\,\hat{P}_{N'}$ 
for an operator $\hat{\cal O}$, with $\hat{P}_{N}$ 
a~projection operator onto the $[N]$ boson space. 
Similarly, the E2 operator is defined as,
\ba
\hat{T}(E2) = e_{B}^{(N)}\,\hat{Q}^{(N)} + e_{B}^{(N+2)}\,\hat{Q}^{(N+2)} ~,
\label{T-E2}
\ea
with boson effective charges, $e_{B}^{(N)}$ and $e_{B}^{(N+2)}$. 

The experimental energies and E2 rates 
for both normal and intruder levels in $^{110}$Cd, are shown in Fig.~2(a).
They are well reproduced by an IBM-PDS-CM calculation, shown in Fig.~2(b),
employing the Hamiltonian of Eq.~(\ref{Hfull}) and the E2 operator 
of Eq.~(\ref{T-E2}). The mixing between the intruder states 
and normal states of class~A is weak. The latter states retain a high 
degree of purity and good U(5) quantum numbers. This is reflected in 
their E2 decay properties, which are essentially the same as those 
of the U(5)-DS shown in Fig.~1(b). 
In contrast, the states in classes~B and C, 
whose decay properties show marked deviations from the U(5)-DS limit,
are mixed with other normal and intruder states. 
The resulting calculated values:
$B(E2;\,0^{+}_3\!\rightarrow\! 2^{+}_1) \!=\! 0.25$, 
$B(E2;\,2^{+}_5\!\rightarrow\! 4^{+}_1) \!=\! 0.19$,
$B(E2;\,2^{+}_5\!\rightarrow\! 2^{+}_2) \!=\! 0.12$ W.u.,
are consistent with the measured upper limits:
$7.9,\,5,\,0.7^{+0.5}_{-0.6}$ W.u., respectively. 

In summary, we have considered the vibrational structure of $^{110}$Cd, 
by means of U(5)-PDS.
The PDS Hamiltonian retains good U(5) 
symmetry for yrast states, but breaks it in particular non-yrast states. 
The mixing with the intruder levels is weak, and affects mainly the broken 
U(5)-DS states. Most low-lying normal states maintain the 
vibrational character and only specific states 
exhibit a departure from this behaviour, in line with the empirical data. 
Calculations are underway to see if this approach can be implemented in other 
neutron-rich Cd isotopes.

This work 
is supported in part (A.L. and N.G.) by the Israel Science 
Foundation (Grant 586/16).


\end{document}